# Enabling microbiome research on personal devices

Igor Sfiligoi
University of California San Diego
La Jolla, CA, USA
isfiligoi@sdsc.edu

Daniel McDonald
University of California San Diego
La Jolla, CA, USA
danielmcdonald@ucsd.edu

Rob Knight
University of California San Diego
La Jolla, CA, USA
robknight@ucsd.edu

*Abstract*—Microbiome studies have recently transitioned from experimental designs with a few hundred samples to designs spanning tens of thousands of samples. Modern studies such as the Earth Microbiome Project (EMP) afford the statistics crucial for untangling the many factors that influence microbial community composition. Analyzing those data used to require access to a compute cluster, making it both expensive and inconvenient. We show that recent improvements in both hardware and software now allow to compute key bioinformatics tasks on EMP-sized data in minutes using a gaming-class laptop, enabling much faster and broader microbiome science insights.

*Keywords—microbiome, bioinformatics, benchmarking, gpu*

## I. Introduction

Recent improvements in DNA sequencing techniques have resulted in drastic acceleration of microbiome research. As experimental designs transition from hundred samples to tens of thousands of samples, so does the associated computational cost in the analysis pipeline. After collecting and sequencing samples, one of the first questions researchers ask is how similar those samples are to each other, and can these similarities be explained by study variables (e.g., salinity, pH, host or not host associated, etc). The computation of sample similarity is performed pairwise producing a distance matrix of size N x N where N is the number of samples being evaluated. Note that researchers also frequently turn to data reuse, by placing new sample collections into the context of existing data, which rapidly increases the number of samples being compared.

Following the construction of a distance matrix, researchers often apply principal coordinates analysis (PCoA), a dimensionality reduction technique similar to principal components analysis, for visualization and statistical purposes. In addition, researchers also frequently wish to ask whether one distance matrix is correlated to another through the use of a non-parametric Mantel test; there are many ways to construct a distance matrix, and many ways to assess the composition of a microbiome sample.

A popular microbiome analysis package is QIIME, which in turn relies on scikit-bio and unifrac packages for core parts of the computation. Until recently, analyzing any sizable number of samples using these tools was considered too slow to perform on personal devices, such as personal workstatioons and laptops. For example, just computing the UniFrac distance matrix of the 25k samples of the Earth Microbiome Project (EMP) [1] would have taken over 8 hours on a server-class 16-core Intel Xeon Gold 6242 CPU.

This work was partially funded by the US National Research Foundation (NSF) under grants DBI-2038509, OAC-1826967, OAC-1541349 and CNS-1730158.

Pre-print, July 2021 version

The authors of this paper thus dedicated effort to do a performance analysis of the key libraries involved and optimized several of them [2,3]. The net result of the work was reducing the run times needed by several orders of magnitude, making many microbiome analyses semi-interactive even on a personal device, such as a gaming-class laptop.

## II. Computing the Unifrac distance

UniFrac is a phylogenetic measure of beta-diversity that assesses differences between pairs of microbiome profiles. UniFrac is useful for microbial community analysis because it can account for the evolutionary relationships between microbes present within a sample. Other distance metrics, such as Euclidean distance, Bray-Curtis, and Jaccard, make the unrealistic implicit assumption that all organisms are equally related, leading to statistical artifacts, particularly with sparse data matrices that are typical in real-world cases as most kinds of microbes are not found in most locations.

A highly parallelizable algorithm, named Striped UniFrac was proposed and implemented a couple of years ago. After adding GPU offloading [2], fp32 compute support and sparse-aware optimizations, the new algorithm, which we call Hybrid UniFrac, is now fast enough to compute the Unifrac distance on up to 50k samples for both unweighted and the weighted normalized versions, especially when using a GPU, as shown in Fig. 1 and Fig. 2. On CPU-only system, going beyond EMP-sized datasets, i.e. around 25k samples, is still a bit slow.

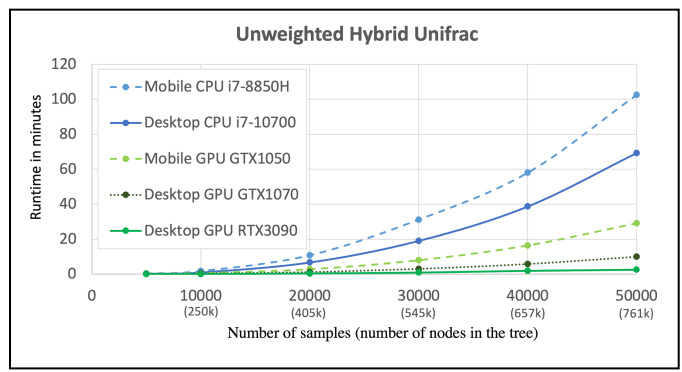

Fig. 1.  Measured runtimes of the unweighted variant of Hybrid Unifrac.

Memory consumption could potentially be a problem, especially when dealing with lower-end GPUs, like the tested NVIDIA GTX1050 mobile GPU, which comes equipped with only 2GB of GPU-addressable memory. To account for that, the provided implementation can split the problem in smaller pieces, i.e. a set of stripes. Those pieces do have to be eventually merged, and that step cannot be further split, but even at 50k samples it requires only 10 GB of RAM and can be completed in 4 minutes on an Intel Core i7-8565U mobile CPU. Moreover, PCoA can also be computed as part of the Hybrid UniFrac compute itself. This implementation has been optimized to compute the PCoA values with virtually no additional memory usage. The needed memory for computing PCoA of a 50k x 50k resulting matrix requires less than 10 GB of RAM.

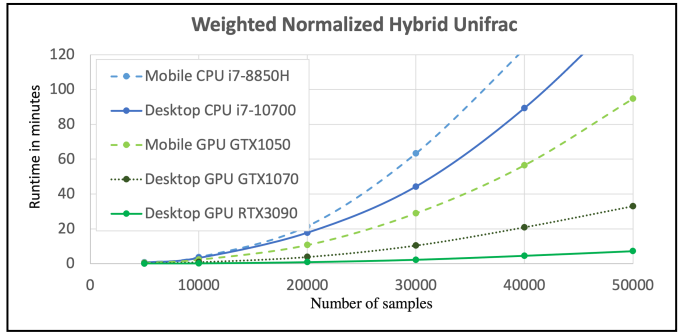

Fig. 2. Measured runtimes of the weighted normalized variant of Hybrid Unifrac.

### III. THE MANTEL TEST

The Mantel test is often used to check for correlations between two distance matrices; there are many ways to construct a distance matrix, and many ways to assess the composition of a microbiome sample. Correlation can be computed using either Pearson's product-moment correlation coefficient or Spearman's rank correlation coefficient.

The latest implementation available in the scikit-bio library [3], which has been re-written in Cython, is now fast enough to compute the Mantel tests on EMP-sized datasets on most personal devices, even mid-range laptops, as seen in Fig. 3. On an Intel Core i7-8565U mobile CPU it is now possible to compute Pearson correlation of 25k x 25k matrices in about 3 minutes, while the Spearman variety requires about 18 minutes.

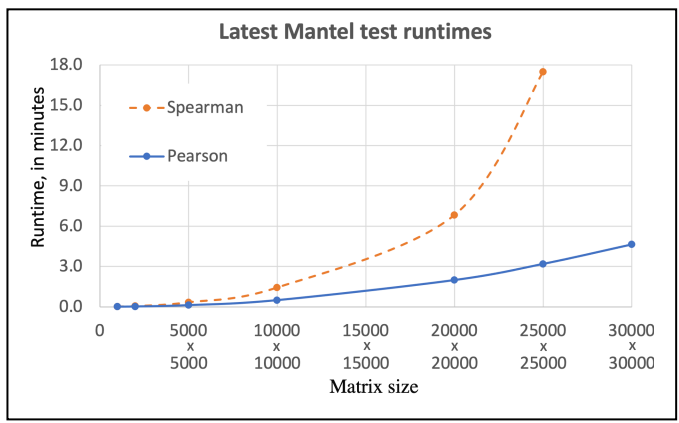

Fig. 3. Measured runtimes for computing a Mantel test using the latest implementation on an Intel Core i7-8565U mobile CPU.

Memory utilization may however still be a problem. Due to the fact that we operate on several matrices at once, at 25k x 25k we already require about 10 GB of RAM for the Pearson version and 24 GB of RAM for the Spearman version, as shown in Fig. 4. Swap space may help a little bit, and we were able to compute the Spearman version of 25k x 25k matrices on a Linux-based laptop with only 16 GB of physical RAM.

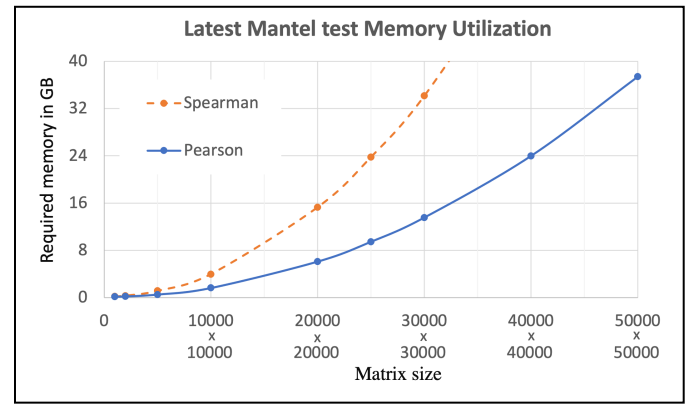

Fig. 4. Measured peak memory use for computing a Mantel test.

The memory usage limited our tests on the desktop-class Intel Core i7-10700 CPU-based system, which had only 10 GB of avaialble RAM. This effectively restricted the tests to 25k x 25k for the Pearson variant and 10k x 10k for the Spearman variant. Although the sizes were relatively small, the runtimes were surprisingly slightly longer on the desktop system compared to the tested mobile system, indicating that the application is likely mostly memory bound. Finally, we want to point out that that we opted not to pursue a port to GPUs. Given both acceptable run times on CPUs and small GPU memory sizes which would severely restrict the computable matrix sizes, it was deemed not worth the effort.

### IV. SUMMARY

We show that it is possible to compute the PCoA of a UniFrac distance of up to 50,000 samples in just minutes using gaming-class, GPU-equipped personal devices, both desktops and laptops. Furthermore, researchers that wish to ask whether one distance matrix is correlated to another through the use of a non-parametric Mantel test can compute the Pearson variant of the Mantel test for up to 30,0000 samples in a few minutes on any mainstream device with at least 16 GB of RAM.

### ACKNOWLEDGMENT

This work was partially funded by the US National Research Foundation (NSF) under grants DBI-2038509, OAC-1826967, OAC-1541349 and CNS-1730158.